\documentclass[copyright]{eptcs}
\providecommand{\event}{PDMC 2009} 
\providecommand{\volume}{??}
\providecommand{\anno}{2009}
\providecommand{\firstpage}{1}
\providecommand{\eid}{??}

\usepackage{graphicx}
\usepackage{amsmath,amssymb}
\usepackage{multirow}
\usepackage{subfigure}

\usepackage{listings}
\lstnewenvironment{code}
    {\lstset{}%
      \csname lst@SetFirstLabel\endcsname}
    {\csname lst@SaveFirstLabel\endcsname}
\newcommand{\BlenX}{BlenX}

\newcommand{\comment}[1]{}
\newcommand{\EQ}{=}

\newcommand{\IN}{\in}

\newcommand{\ignore}[1]{}

\title{Efficient Parallel Statistical Model Checking of Biochemical Networks}
\author{P. Ballarini \qquad M. Forlin \qquad T. Mazza \qquad D. Prandi
\institute{The Microsoft Research - University of Trento Centre for Computational and Systems~Biology}
\email{\{ballarini,forlin,mazza,prandi\}@cosbi.eu}
}
\def\titlerunning{Efficient Parallel Statistical Model Checking of Biochemical Networks}
\def\authorrunning{Ballarini, Forlin, Mazza  \& Prandi}
\begin{document}
\maketitle

\begin{abstract}
  We consider the problem of verifying stochastic models of biochemical networks 
  against behavioral properties expressed in temporal logic terms. 
  Exact probabilistic verification approaches 
  such as, for example, CSL/PCTL model checking, are undermined by 
  a huge computational demand which rule them out for most real case studies. 
  Less demanding approaches, such as statistical model checking, estimate the likelihood  
  that a property is satisfied   by sampling   \emph{executions} 
  out of the stochastic model.  
  We propose a methodology for efficiently estimating the likelihood that a LTL property $\phi$ holds of a  
  stochastic model of a biochemical network. As with other statistical verification techniques, the methodology we propose 
  uses a stochastic simulation algorithm for generating execution samples, however there are three key aspects that improve the efficiency: 
  first, the sample generation is driven by on-the-fly verification of $\phi$ which results in optimal 
  overall simulation time. 
  Second, the confidence interval estimation for the probability of $\phi$ to hold is based on an efficient variant of 
  the Wilson method which ensures a faster convergence. Third, the whole methodology is   designed according to a parallel fashion and a prototype 
  software tool has been implemented that performs 
  the  sampling/verification process in parallel over an HPC architecture. 
\end{abstract}

%
%
\section{Introduction}
\label{sec:Intro}
Systems biology~\cite{Kitano02} is concerned with developing detailed models of complex biological networks, which 
then need to be validated and analysed. 
A model of a biological system essentially describes 
the dynamics of a population of $n$  interacting biochemical species $S_1, S_2, \ldots S_n$. 
Analysing the behaviour of such  models entails looking for the 
occurrence of biologically relevant events during the evolution of the system, 
a problem whose complexity is exponentially proportional to the population 
of the considered model.

In this paper we consider 
discrete stochastic modelling of biological systems.  In the discrete-stochastic setting, 
biochemical species are enumerable quantities representing 
the number of molecules of a given substance, and the evolution of the 
system is probabilistic, rather than 
deterministic, leading to  
Continuous Time Markov Chain (CTMC) models. 

Classical \emph{transient-state} and \emph{steady-state} analysis~\cite{Stewart94}
allows to understand important features of a CTMC model. 
Unfortunately numerical solution of Markov models 
is undermined by storage requirements which explode with the dimension 
of the model. Although techniques for the efficient storing of CTMC's matrix/vector 
have been widely studied (see for example~\cite{MiCi99,ScBo98}), solving  
CTMC models   remains unfeasible, in most realistic case studies. 

In recent times, query based verification methods (i.e. model checking~\cite{Clarke99}) 
proved to be valuable instruments for a more expressive analysis of 
stochastic biological models~\cite{KNP09e,ballarini2009studying,Barnat:1211142}. 
Biologically relevant events can be  formally characterised as temporal logic formulae that can then 
be automatically checked against a discrete state model. 
\emph{Exact} probabilistic PCTL~\cite{HansJons89} and stochastic CSL~\cite{BHHK03,aziz00} model checking suffer, as well,  of the state-space explosion problem  which limits their accessibility specially in systems biology where very large models are just common.
\emph{Statistical model checking}~\cite{YKNP06,Clarke08,Donaldson08} has been proposed as an alternative to exact probabilistic model checking that 
allows for getting an estimate  of the likelihood of a condition to hold of a CTMC model. 
 
The \emph{statistical model checking} paradigm essentially is  a 
combination of  efficient stochastic simulation with model checking procedure. It comprises three ingredients: 
(\emph{i}) a stochastic engine that generates trajectories of the underlying state space; (\emph{ii}) a model checking
algorithm capable of analyzing a single trajectory; (\emph{iii}) a statistical support for estimating 
the accuracy of the answer. The advantage of statistical model checking is that, contrary to 
exact model checking, it does not 
need to build (i.e. store) the state space of the model as it only explores a limited number of (finite) 
trajectories. The cost paid for such space saving is in terms of precision of the calculated measure. 

We propose a methodology that given a CTMC model $M$, a property $\phi$ and a 
desired \emph{level of confidence}  estimates the  probability of $\phi$ to be satisfied by $M$ 
with  the estimated measure meeting the desired confidence. 
The method we propose is based on three key aspects. 
Firstly  the trajectory generation  is controlled  by on-the-fly verification of the considered 
formula which means that simulation halts as soon as a state which verifies (falsifies) the formula is reached.
Secondly,  we use an efficient  statistical method 
(i.e. a  variant of the  \emph{Wilson score interval method}) which results in   
smaller  samples  (i.e fewer simulation runs) in order to 
meet the desired confidence. 
Finally the whole simulation/verification framework has been designed and tested
on a parallel prototype, which is based on independent simulation/verification engines 
generation, and a MPI client/server parallel computation architecture.  

The reminder of the paper is organised as follows: in Section~\ref{sec:BioNetworks} 
we introduce the basics about CTMC modelling and 
stochastic simulation algorithms. 
In Section~\ref{sec:statLTLMC} we describe the core of our statistical model checking 
procedure; in Section~\ref{sec:tool} the architecture of the prototype 
which exploits the methodology is described. In Section~\ref{sec:Experiments} we provide evidence of application of statistical verification 
of relevant properties to an abstract model of the budding yeast cell cycle. 
Concluding remarks are given in the final section of the paper.

\section{Stochastic Simulation of Biochemical Networks}
\label{sec:BioNetworks}
The time evolution of a well stirred biochemical reacting system can be described as a 
Continuous-time Markov Chain (CTMC) whose states are vectors $\vec{X}(t)$ of discrete random
variables $X_i(t)$, that give the amount of a molecule $i$ at time $t$. The associated joint probability distribution 
$\mathcal{P}(\vec{X},t)$, often called \emph{Chemical Master Equation} (CME), gives a precise description of 
a biochemical network~\cite{gillespie1992rigorous}. 
Unfortunately, as soon as the system becomes non-trivial,
it is nearly impossible to solve CME (i.e. the corresponding CTMC), neither analytically nor numerically. 
The Stochastic Simulation Algorithm~\cite{Gillespie77} is a computer program that takes a biological reacting system and produces a trace in the space of the CME. 
In the following we first introduce the basics about CTMCs and then we describe  
how the stochastic simulation algorithm works. 

\subsection{Continuous Time Markov Chains}
\label{sec:CTMC}
A Continuous-time Markov Chain (CTMC) for biological modelling is a tuple $M=(S,s_0,Q,Var,Eval)$ 
where $S$ is a finite set of states, $s_0$ is the initial state, $Q$ is the \emph{infinitesimal generator 
matrix} of the chain, $Var$ is the set of \emph{state-variables} and $Eval:S\times Var \to \mathbb{R}_{\geq 0}$ is the \emph{evaluation function} that, given the variable $x$ and state $s$, returns the value of $x$ in $s$. 

A path of a CTMC $M$ is a possibly infinite sequence of $S\times\mathbb{R}_{> 0}$ pairs $\sigma\equiv (s_0,t_0), (s_1,t_1) \ldots (s_n,t_n)\ldots$ describing a trajectory of $M$, also denoted:
$$
s_0\stackrel{t_0}\longrightarrow s_1\stackrel{t_1}\longrightarrow s_2\ldots\stackrel{t_{n-1}}\longrightarrow s_n\stackrel{t_n}\longrightarrow\ldots
$$
where $\forall i\in\mathbb{N}$  $s_i\in S$ and $Q(s_i,s_{i+1})>0$ 
and $t_i\in\mathbb{R}_{>0}$. 
A point $(s_i,t_i)$ of   $\sigma$ indicates that at the $i$-th step of the execution 
$\sigma$ the system was in state $s_i$ and that it stayed in $s_i$ for $t_i$. 
We adopt the following notation: $\sigma[i]$ denotes 
the $i$-th state of $\sigma$; $\sigma^i$ denotes 
the $i$-th suffix of $\sigma$; $\delta(\sigma, i)$ denotes the time spent in the $i$-th 
state of $\sigma$ whereas $\sigma@ t$ denotes which state $\sigma$ is in at time $t$. 

The state-variables $Var$ of a CTMC \emph{biochemical} model will 
coincide with  $ \{ S_1, \ldots, S_N \}$, the set of biochemical species. 

\subsection{Stochastic Simulation Algorithm}
\label{sec:SSA}
The Stochastic Simulation Algorithm (SSA) supposes a well-stirred set $ \{ S_1, \ldots, S_N \}$ of
biochemical species reacting through $M \geq 1$ reaction channels (reactions for short) $ \{ R_1, \ldots, R_M \}$. A
reaction channel $R_j$ is characterized by:
\begin{itemize}
\item the probability $a_j(\vec{x}) dt$ that, given $\vec{X}(t) = \vec{x}$, one reaction $R_j$ will occur in the
next infinitesimal interval $[ t, t + dt )$;
\item the change $v_{ji}$ of the number of molecules of the specie $S_i$ produced or consumed by a reaction $R_j$.
\end{itemize}
The physical and mathematical explanation of $a_j(\vec{x}) dt$ is described in \cite{gillespie1992rigorous}. Basically,
$a_j(\vec{x})$ is a function of the number of possible active instances of reaction $R_j$. Consider, for example, a reaction 
$R_1: S_1 + S_2 \rightarrow S_3$, then $a_1(\vec{x}) = c_1 x_1 x_2 $, where $x_1 x_2$ is the number of active $R_1$ in the
current state $\vec{x}$, and $c_1$ is a constant that depends on the physical characteristics of $S_1$ and $S_2$. 

Given  $a_j(\vec{x})$, the evolution of a biochemical network is described by the 
\emph{next reaction density function} $p(\tau, j | \vec{x}, t )$, i.e.
the probability, given $\vec{X}(t) = \vec{x}$, that the next reaction in the system will occur in the infinitesimal time 
interval $[ t + \tau, t + \tau + dt )$ and will be on channel $R_j$:
\begin{equation}
p(\tau, j | \vec{x}, t ) = a_j( \vec{x} ) e^{ - a_0(\vec{x}) \tau } \label{eq:nextreaction}
\end{equation}
where $a_0 = \sum_{j=1}^{M} a_j(\vec{x})$. By conditional probability, Eq.~\ref{eq:nextreaction} becomes
$p(\tau, j | \vec{x}, t ) = p_1(\tau | \vec{x}, t)p_2(j | \tau, \vec{x}, t)$ where
\begin{subequations}
\begin{equation}
p_1(\tau | \vec{x}, t) = a_0(\vec{x}) e^{ - a_0(\vec{x}) \tau }   \qquad\ \qquad (\tau \geq 0)   \label{eq:time}
\end{equation}
\vspace{-3ex}
\begin{equation}
p_2(j | \tau, \vec{x}, t) = \frac{a_j(\vec{x})}{a_0(\vec{x})}    \label{eq:reaction}
\end{equation}  
\end{subequations}
meaning that, $\tau$ is a sample from an exponential random variable with rate $a_0({\vec{x}})$,
and the selected reaction $j$ is independently taken from a discrete random variable with values in
[1,M] and probabilities $\frac{a_j(\vec{x})}{a_0(\vec{x})}$.

Standard Monte Carlo methods are then used to select time consumed and next reaction according to
Eq.~\ref{eq:time} and Eq.~\ref{eq:reaction}, to produce
a trajectory in the discrete state-space of the CME. 
Several variants of the SSA exist~\cite{li2008algorithms} that differ 
in how the next reaction is selected and in the data structures used, 
but all of them are based on the following common template: 
\begin{description}
\item[1]	Compute $a_0$
\item[2]	Randomly select a reaction $j$ according to Equation~\ref{eq:reaction}
\item[3]	Randomly select a time duration $\tau$ according to Equation~\ref{eq:time}
\item[4]	Update state vector as $\vec{x} \gets \vec{x} + v_{j}$ and total time as $t \gets t + \tau$
\item[5]	Go to step 1 or Terminate
\end{description}

\section{On-the-fly statistical BLTLc Model Checking}
\label{sec:statLTLMC}
Statistical verification of a CTMC model $M$ is based on the simple principle of 
collecting $N$ sample realisations $\sigma_i$ ($i\in\{1,\ldots ,N\}$) 
of  $M$ and verifying each of them against a given property $\phi$. 
The estimate of the likelihood of $\phi$ to hold true of $M$ is 
obtained as the frequency $\hat{p}_{\phi}=\frac{po}{N}$ of positive outcomes ($po$)  of the verification of $\phi$ versus $\sigma_i$. 
In the following we introduce the temporal logic we refer to as the language for stating properties of simulated trajectories, namely the BLTLc logic.  We then provide details of  a procedure for on-the-fly  simulation-verification of a CTMC model which combines the random generation of a trajectory,  by means of stochastic simulation (see Section~\ref{sec:BioNetworks}), with a verification procedure 
based on the BLTLc semantics. Finally we describe 
a statistical procedure for the efficient estimation of the likelihood of a  formula. 
This procedure iteratively works  out the number of simulation runs needed in order to meet the desired  level of confidence.

\subsection{Bounded Linear-time Temporal Logic with numerical constraints}
\label{sec:LTL}
We introduce the  Bounded Linear-time Temporal Logic with numerical constraints 
(BLTLc) a logic  for stating properties referred to timed-trajectories resulting from 
stochastic simulation of a CTMC model. 
BLTLc combines  the  Constraint-LTL~\cite{Fages06,DBLP:conf/cmsb/FagesR07} ($LTLc$ or $LTL(\mathbb{R})$) for \emph{arithmetic-constrained} LTL formulae with  the Bounded-LTL (BLTL)~\cite{Clarke08} for time-bounded  LTL formulae. 
Both LTLc and BLTL are based on classical LTL~\cite{Pnueli77} temporal operators.  
However while LTLc allows for using (complex) arithmetical conditions 
between state variables, it does not allow for expressing time bounded conditions. On 
the other hand with BLTL time bounded LTL expressions can be formed 
but based on simple non-arithmetical conditions  rather 
than  on a grammar for arithmetic expressions as it is the case 
with LTLc. 
The syntax of the BLTLc logic is given by the following grammar:
\begin{align*}
\phi &:= val \trianglelefteq val \mid \lnot\phi  \mid\ \phi\lor\phi \mid\ \phi \land\phi \mid X^I\ \phi\mid  \phi\ U^I\phi\\ \\
val &:=  x  \mid val \sim val \mid func \mid Int \mid Real
\end{align*}
 \noindent where $I\subseteq\mathbb{R}_{\geq 0}$, $\trianglelefteq\IN\{<,\leq,\geq,>,=,\neq\}$, 
$\sim\IN\{+,-,*,/\}$ and $\mathit{func}$ denotes 
common functions such as $pow(),sqrt(),\ldots $. 
Note that unbounded formulae correspond to $I\EQ [0,+\infty)$ (for simplicity $ [0,+\infty)$
is usually omitted for unbounded formulae). 
The operators $\lnot$, $\lor$ and $\land$ are the standard boolean logic connectives \emph{not}, 
\emph{or} and \emph{and} respectively, whereas $X^I$ and $U^I$ denote the \emph{temporal connectives}, \emph{next} and \emph{until}, respectively. 
The \emph{next} operator refers to the notion of \emph{true in the next state} within time 
$I$ whereas the  \emph{until} operator ($\phi\ U^I \psi$) indicates that a future state where  its second argument  ($\psi$) holds is reached within time $I$ and while its first argument ($\phi$) continuously holds.  The two popular  \emph{Finally} 
operator ($F^I (\phi)$, which refers to the notion of a condition holding true at some point in the future 
and within time $I$), and \emph{Globally} 
operator ($G^I (\phi)$, which refers to the notion of a condition to continuously  holding true 
within time $I$), are also supported by BLTLc relying on  the well-know  equivalences 
 $F^I(\phi)\equiv[ tt\ U^I \phi]$ and $G^I(\phi)\equiv\neg F^I(\neg\phi)$.

BLTLc formulae are evaluated against timed-paths resulting from simulation of a CTMC model. 
The formal semantics of BLTLc formulae, expressed in terms of the $\models$ relation, is given below, 
where $\sigma$ is a timed-path of a CTMC model.
\begin{itemize}
\item $\sigma\models (val'\trianglelefteq val'')$ if and only if $Eval(\sigma[0],val')\trianglelefteq Eval(\sigma[0],val'')$
\item $\sigma\models \lnot \phi $ if and only if $\sigma\not\models\phi$
\item $\sigma\models \phi'\lor\phi'' $ if and only if $\sigma\models\phi'$ or $\sigma\models\phi''$
\item $\sigma\models \phi'\land\phi'' $ if and only if $\sigma\models\phi'$ and $\sigma\models\phi''$
\item $\sigma\models X^I \phi $ if and only if $\sigma^1\models\phi$ and $\delta(\sigma,0)\IN I$
\item $\sigma \models \phi' \ U^I\ \phi''$ if and only if $\exists i\IN\mathbb{N}:  
  \sigma^i\models\phi''$ and  $\sum_{j<i} \delta(\sigma,j) \IN I$ and $\forall j<i$, $ \sigma^j\models\phi' $
\end{itemize}
$Eval(\sigma[i],val)$ is the \emph{evaluation function} that assigns a numerical value to 
the expression $val$ by looking up at the value that each variable $x\IN Var(val)$  has in 
state $\sigma[i]$ of path $\sigma$ (see Section~\ref{sec:CTMC}).

As  an example of the expressiveness of the BLTLc logic consider the following formula  $\phi\equiv[(X_1<Sqrt(X_2))\ U\ (X_2\geq 10+X_3)]$ which states that the concentration of $X_1$ shall be less than the square root of that of $X_2$ until that of $X_2$ exceeds 
$X_3$ by at least 10.
 
\begin{figure}[t!]
\hrulefill 
\begin{lstlisting}
// The procedure returns a (Result,Trace) pair, where Result is the boolean 
// result of verification and Trace is the simulation trace generated.

(Result, Trace) simulate_verify(i, sigmabuff, M ,phi, tmax)
  (s,t) = sigmabuff[i];
  case phi of
    propositional formula: return (Eval(psi, M, s), sigmabuff);
    neg psi: (r, tr) = simulate_verify(i, sigmabuff, M, phi, tmax);
         return (not r, tr);
    psi1 and psi2: (r, tr) = simulate_verify(i, sigmabuff, M, phi, tmax);
            (rprime, trprime) = simulate_verify(i, tr, M, phi, tmax);
            return (r and rprime, trprime);
    nextI psi: if (t notin I) then return (false, sigmabuff);
          if (i == max_index(sigmabuff))
             { (snext, tnext) = Next((s,t), M);
               sigmabuff += (snext, tnext); }
          return simulate_verify(i+1, sigmabuff, M, phi, tmax);
    psi1 untilI psi2: if (t notin I) return (false, sigmabuff);
            (r, tr) = simulate_verify(i, sigmabuff, M, psi2, tmax);
            if ( last(tr).time > tmax) return (false, tr);
            if ( r ) then return (true, tr);
            (rprime, trprime) = simulate_verify(i, tr, M, psi1, tmax);
            if ( neg rprime ) return (false, trprime);
            if (i == max_index(trprime))
               { (snext, tnext) = Next((s,t), M);
                 sigmabuff += (snext, tnext); }
            return simulate_verify(i+1, trprime, M, phi, tmax);             
\end{lstlisting}
\hrulefill
\caption{On-the-fly verification of simulated trajectory}
\label{fig:verification}
\end{figure}
Finally we stress that differently from the original BLTL~\cite{Clarke08},  the association of {probabilistic bounds} 
to BLTLc formulae is not supported. This is because 
the aim of the statistical verification procedure we introduce is to estimate the 
actual measure of probability of a BLTLc  formula rather than to estimate 
whether such measure is below/above a certain threshold. 

\subsection{On-the-fly verification of simulated trajectory}
\label{sec:onthefly}
We define a procedure that given a CTMC model $M$ and a BLTLc formula $\phi$ 
allows for stochastically generating trajectories of $M$ while verifying whether the 
generated trajectories satisfy or falsify the considered formula $\phi$. 
The verification of $\phi$ is performed \emph{on-the-fly} 
meaning that  simulation proceeds with the generation of the next state 
 only if $\phi$ is neither satisfied nor falsified in the current one 
(and if the simulation time limit has not been reached). 
The pseudocode for the verification procedure is  outlined in 
Figure~\ref{fig:verification}. Function \emph{simulate\_verify()} takes few input parameters:  
 a BLTLc formula $\phi$, a  trace $\sigma_{buff}$ (containing the 
 \emph{pre-viewed} trace resulting from recursive calls corresponding to the verification of sub-formulae)
the current position $i$ in the buffered trace, a CTMC $M$ and a maximum simulation time $t_{max}$.   
The algorithm works in the following manner: 
propositional formulae are verified trivially relying on the function
\textit{Eval}$(\phi, M, s)$, that gives the value of a non temporal formula $\phi$ in a state $s$. 
On the other hand temporal formulae (may) require the generation of a simulation-trace, 
obtained through a call to the the stochastic engine  function \textit{Next}$((s,t),M)$. Verification of temporal formulae may result in passing of the 
already-generated trace (i.e. the buffered trace $\sigma_{buff}$ that results from verification of 
a sub-formula) from the inner-most sub-formulae to the outer-most ones. This 
is required for two reasons: first to avoid possible branching during the generation of a single 
trace (if verification requires to roll-back to a previously generated state then re-application 
of the SSA may result in the generation of a different successor state, else in branching, which 
is wrong in the LTL context), secondly 
for efficiency (the already generated traces should be re-used whenever feasible). 
For these reasons a call to the \emph{simulate\_verify()} returns a pair $(r,tr)$ where $r$ is the boolean result of the verification of the considered formula $\phi$ and $tr$ is the simulation trace generated up until 
verification of $\phi$. The buffer trace $\sigma_{buff}$ in the initial call will contain a single element  $\sigma_{buff}=\{(s_0,t_0)\}$, representing the initial state and time of simulation. 
Finally, for time-unbounded formulae, we adopt the following \emph{pessimistic} approach: if the simulation 
max time $t_{max}$ is reached and the formula is neither verified nor falsified then the algorithm returns 
$\mathit{false}$. Thus the exact probability of time-unbounded formulae is an upper bound of the 
estimated one.

\subsection{Estimating the probability of a property}
\label{sec:EstimateProb}
Checking of a BLTLc property on a simulated trace corresponds to a Bernoulli experiment, where the outcome can be either positive or negative.
Thus  the number of successes  that results from reiterated checking of the same property on  $n$  independent simulations, represents a random variable $X$  with a Binomial distribution. 
The point estimation of the unknown probability of success $p$ out of $n$ independent trials, is given by the well known maximum likelihood estimator $\hat{p}=po/n$, where $po$ represents the number of successes. Clearly the reliability of such estimate  is highly affected by the number of simulations performed, i.e. by the sample size  $n$. As a consequence 
the point estimate  $\hat{p}$ is usually associated with  a confidence interval, expressed 
in terms of a real value  $\alpha\IN (0,1)$, which represents the range within which
the actual value of the unknown parameter $\theta$ (i.e. the actual probability of the considered formula to hold against the simulated model) shall fall $(1-\alpha)\%$ times\footnote{There  exists a strong connection between confidence intervals and hypothesis testing: all the values $\theta_0$ for the unknown parameter $\theta$ external to a $1-\alpha$ confidence interval would end in the rejection of the two sided hypothesis testing (i.e. Null Hypothesis $H_0: \theta=\theta_0$) at the $\alpha$ level.}. 

The standard approach to compute the confidence interval for the probability of success of a binomial distribution uses  the normal approximation, producing the so-called  Wald interval. 
The hypothesis-testing-based  statistical model checking approach of Younes \emph{et al.}~\cite{YKNP06}, for example, uses the Wald interval method for validating the hypothesis that 
the probability of a certain CSL formula is below (above) a given threshold. 

Brown \emph{et al.} \cite{Brown01intervalestimation},\cite{Brown02confidenceintervals}, have studied the coverage characteristics of different types of binomial proportion confidence intervals, and they showed that the Wald interval present unstable coverage characteristics also for large $n$, suggesting thus the use of other types of confidence intervals. 
Among the interval discussed, the Wilson score interval \cite{wilson27} have shown good coverage characteristics also for small $n$ and extreme probabilities. Wilson score confidence interval is calculated by means of 
$Wilson\_interval(\hat{p},n,\alpha)=[L ,U]$ with,
\begin{equation}
	\label{eq:wilson_limits}
	[L, U]\!=\!\frac{\hat{p}+\frac{1}{2n}z^2_{1-\alpha/2}\mp z_{1-\alpha/2} \sqrt{\frac{\hat{p}(1-\hat{p})}{n}+\frac{z^2_{1-\alpha/2}}{4n^2}}}{1+\frac{1}{n}z^2_{1-\alpha/2}}
	\end{equation}
where $\hat{p}$ is the estimated probability from the statistical sample, $\alpha$ is the confidence level, $z_{1-\alpha/2}$ is the $1-\alpha/2$ percentile of a standard normal distribution, and $n$ is the sample size.
As the confidence interval is a function of $n$ we may reverse the problem and ask which is the proper sample size to obtain a confidence interval of a given width at a specific confidence level $\alpha$. This is extremely useful for  establishing how long 
a simulative experiments has to be (i.e. how many runs are needed) in order to  meet a desired reliability for the measure we estimate. 

\subsubsection{An iterative algorithm for sample size determination}
\label{sec:Wilson}
The sample size required for the Wilson interval of width $2\epsilon$ at $1-\alpha$ confidence level  can be obtained simply by solving for $n$ the Wilson score limits in equation (\ref{eq:wilson_limits}) \cite{Piegorsch2004}: formally 
this is given by function $Wilson\_sample(p,\epsilon,\alpha)=N$ with,
\begin{equation}
	\label{eq:wilson_sample}
	N \geq z^2_{1-\alpha/2}\frac{\hat{p}(1-\hat{p}) - 2\epsilon^2+ \sqrt{\hat{p}^2(1-\hat{p})^2 + 4\epsilon^2(\hat{p}-0.5)^2}}{2\epsilon^2}
\end{equation}
\noindent where $\hat{p}$ is the frequency of positive outcomes of a re-iterated Bernoulli experiment. 
As we usually do not have a guess of the probability $\hat{p}$ to be used in (\ref{eq:wilson_sample}), the standard approach is to take a conservative estimate, by considering $\hat{p}=0.5$ which is the estimate with maximum variance, and, as such, produces the highest sample size.

The method we present here consists in adopting a different approach in determining the sample size. By iterating (\ref{eq:wilson_sample}) with successive estimates of  $\hat{p}$ we are able to drastically reduce the number of samples required when the true probability $p$ is far from 0.5.
More specifically, given a confidence interval width 2$\epsilon$ and a confidence level $1-\alpha$, the algorithm starts by calculating the sample size required for an initial estimate  $\hat{p}=1$ (or equivalently $\hat{p}=0$) and returns the minimum number $N$ of simulations to be performed. After computing the proportion of successes the new estimate $\hat{p}$ is rounded by adding or subtracting the quantity $\epsilon$ if $\hat{p}\leq0.5$ or $\hat{p}>0.5$ respectively. The rounded estimate $p^\prime$ is then used to recalculate the sample size resulting in $N^\prime$. If we have already performed a cumulative $N_{tot}\geq N^\prime$ simulations the algorithm stops. Conversely, we iterate the process again by launching $N^\prime-N_{tot}$ simulations. 

\begin{figure}[htbp]
\hrulefill
\begin{lstlisting}%[numberfirstline=false,numbers=left]
Wilson Procedure
  Ntot = 0;
  N = Wilson_Sample(1, epsilon, alpha);
  Perform N experiments;
  Ntot = Ntot + N; pest = yess;
  if pest <= 0.5
     pprime = pest + epsilon;
  else pprime = pest - epsilon;
  Nprime = Wilson_sample(pprime, epsilon, alpha); Nnew = Nprime - Ntot;
  if Nnew > 0
     N = Nnew; goto 3;
  else return (pest, Wilson_interval(pest, Ntot, alpha));
\end{lstlisting}
\hrulefill
\caption{Iterative Wilson method for sample size determination}
\label{fig:wilson}
\end{figure}
The $\hat{p}$ rounding step is crucial and ensures that a successive sample size calculation would avoid undersized samples due to erratic estimates. If the current estimated probability of success drifts from the true unknown one, towards extreme probabilities, of more than $\epsilon$, this would produce an undersized sample which would produce a confidence interval not covering the parameter at $1-\alpha$ level.   

\subsubsection{Performance of the iterative sample size determination}
The iterative method for the determination of sample size for the Wilson score interval drastically reduces the number of required samples with respect to the conservative approach that starts with $\hat{p}=0.5$. Of course this gain is greater when the actual $p$ is close to extreme values $p=0$ and $p=1$, while using the same sample size for $p$ close to 0.5.

\begin{figure}[htbp] 
   \centering
   \includegraphics[width=4in]{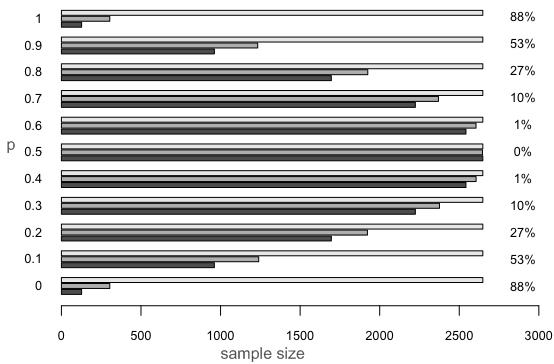} 
   \caption{Average sample size for CI-width$=0.05$ at $99\%$ confidence lelvel. 
   Comparison of sample size required with i) conservative approach, ii) Iterative Wilson, iii) Minimum sample size with known $p$}
   \label{fig:ComparisonSampleSize}
\end{figure} 
Figure \ref{fig:ComparisonSampleSize} represents the sample size required to obtain a confidence interval of a given width $2\epsilon=0.05$ at a confidence level $1-\alpha=0.99$ for values of $p$ ranging from 0 to 1 by 0.1. Bars from light gray to dark gray represent respectively:
\textbf{I}. Conservative approach (using always $\hat{p}$=0.5); \textbf{II}. Iterative Wilson; \textbf{III}. Minimum sample size if the unknown $p$ would be known (i.e. $\hat{p}=p$).
On the right side of the plot we reported the percentage of reduction in sample size by using the Iterative Wilson approach with respect to the Conservative approach.
As it can be easily seen, we have the strongest reduction in sample size when the true $p$ is close to 0 or 1. By using the Iterative Wilson method we presented here, we can reduce the sample size required by up to 88\% with respect to the Conservative approach. This reduction at extreme values is explained by the fact that at those probabilities the variance of the binomial distribution is smaller, so we need less samples to obtain good estimates.

\section{Prototype architecture}\label{sec:tool}
\begin{figure}[t!] 
\begin{center}
   \subfigure[Wilson-driven computational loop]{\includegraphics[width=2.6in]{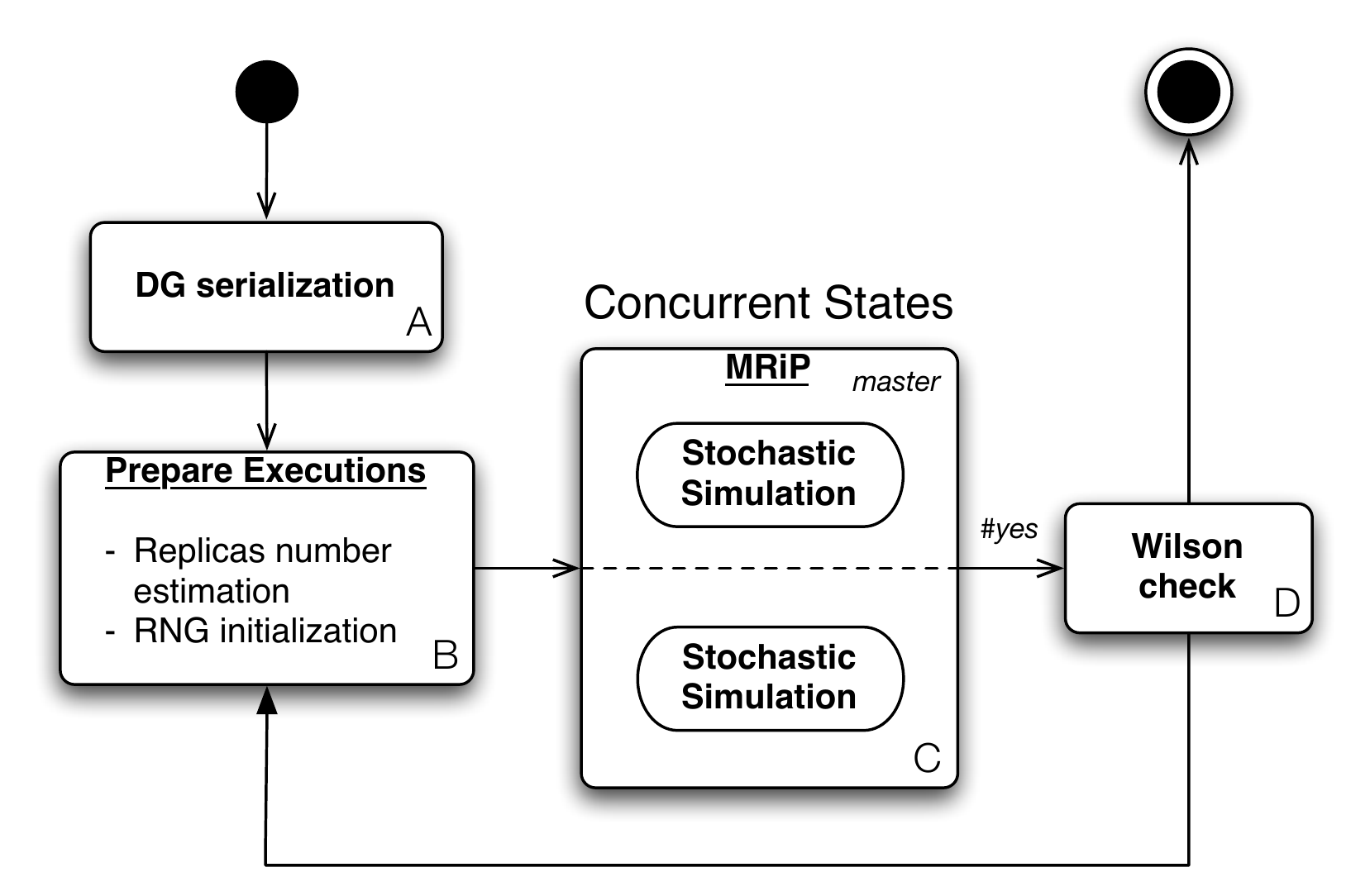}
   \label{fig:statediagram}}
	 \subfigure[Inter-process and client-server communications]{
   \includegraphics[width=2.6in]{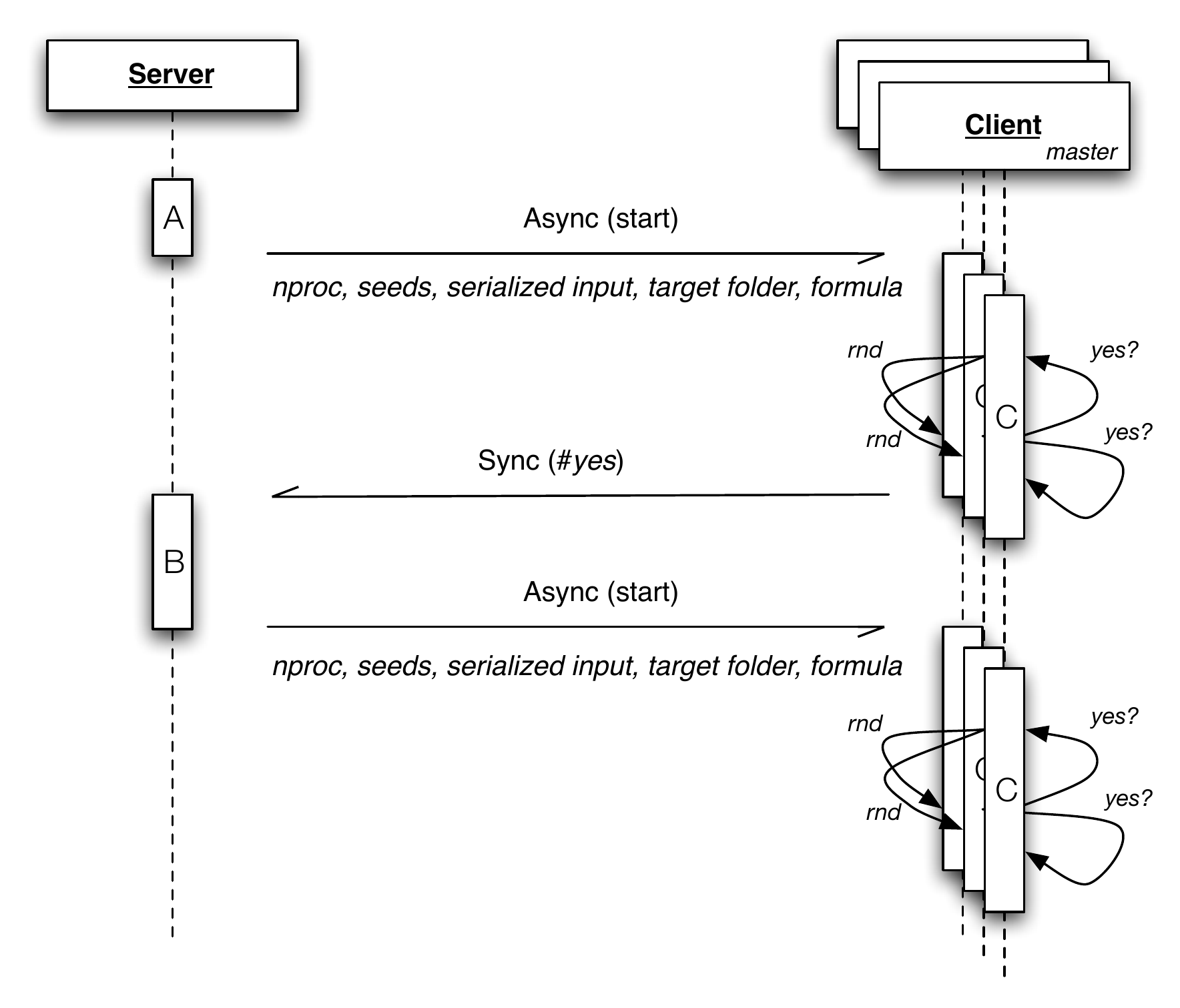} 
   \label{fig:activitydiagram}}
\end{center}
\end{figure}

What so far illustrated has been implemented within a distributed software architecture, capable to get the better out of the MRiP computational policy (refer to \cite{ballarini2009BIB} for a wide overview of this topic). It is actually made of two distinct modules: a graphical front-end and a remote simulation engine. The front-end part acts as a server and is in charge of drawing the computational graph relative to the loaded \BlenX~\cite{dematte2008blenx}
 models. \BlenX\ is a new programming language intended to
develop executable biological models starting from the composition of the description of the molecules involved in the system. 
The prototype
collects any information from a \BlenX\ model and serializes them in a proprietary, xml-based, data format along with all the simulation information manually inputted by the user (see Figure \ref{fig:statediagram} - State \verb|A|). Further information about the logical formula to be checked, the $\alpha$ and $\epsilon$ values are required only whenever one wants to automatically calculate the number of replicated simulations (or ``replicas") needed to reach the required confidence threshold. However, both in the case that the number of replicas is user-defined and that it is automatically computed, a random number generator is instantiated and used to make a stream of initial seeds, one for each simulation (see Figure \ref{fig:statediagram} - State \verb|B|). 

As soon as the simulation task is invoked by the user, a number of independent simulation engines is instantiated (see Figure \ref{fig:activitydiagram} - Activity \verb|A|). Among them, one is entitled to be master. The master handles both the inter-process and the client-server communications. In the former case, it takes care of scattering and dispatching the initial seeds to the slave processes (and to itself) and of gathering the results (see Figure \ref{fig:activitydiagram} - Activities \verb|C|). \verb|MPI| is broadly used for these purposes. In the latter case, the master node is responsible for counting the computed \verb|YES| and for its sending to the server (see Figure \ref{fig:activitydiagram} - Activities \verb|C|). This is accomplished by means of a socket-based interface. Hence, each process simulates independently (see Figure \ref{fig:statediagram} - State \verb|C| and Figure \ref{fig:activitydiagram} - Activities $C$) and evaluates on-the-fly a logical formula, giving a boolean answer. The summation of the positive answers is sent to the server, which recomputes the Wilson method and returns a new number of simulation replicas to be performed (see Figure \ref{fig:statediagram} - State \verb|D| and Figure \ref{fig:activitydiagram} - Activity \verb|B|). This loop halts only when no more replicas are required to be performed. 

\subsection{Policies of communications}
The prototype initializes as many independent processes as the replicas number estimated by the Wilson procedure. In agreement with the SIMD (Single Instruction, Multiple Data) computational policy, they share the same code area, namely they execute the same instructions in a parallel fashion on different instances of the same input model. In particular, two important tasks are executed in parallel: the import of the input model and the actual simulation. The former task consists just in loading a \BlenX\ model and in extracting some important information from it. This can happen essentially because all the processes address a chunk of shared filesystem area where the input model lies in. After that, every process starts its independent simulation according to the parameters entered by the user.

Therefore, we set two synchronization points. One occurs before starting the simulation task, exactly when the starting random seeds are computed. Actually, this would not be strictly necessary because, in principle, every process could generate its own starting random seed. But, to tackle the problem of avoiding correlation among the individual trajectories of random numbers~\cite{tian2005parallel}, we empowered the master process (the process with the lowest \verb|MPI rank|) to generate as many random numbers as the number of processes and to split and distribute it to each other process (through the \verb|MPI_Scatter| procedure). Contrarily, a very critical synchronization point concerns the Wilson step. It occurs whenever all the processes end their simulation and must communicate their \verb|YES| number as well as must write the simulation traces to filesystem. Here, two kind of communications take place. One happens among the processes themselves. Anyone send the master its computed simulation trace along with a boolean answer to the formula (via the \verb|MPI_Gather| procedure), both encapsulated in an ad-hoc serializable data structure. Thus, the master node prints the traces to file and computes the probability $\hat{p}$, as described before. Finally, it sends back $\hat{p}$ to the server side via a private socket for the Wilson re-computation.

\section{Experiments and Performances}
\label{sec:Experiments}
We consider a stochastic model  of (a part) of the regulatory network that controls the  buddying yeast cell cycle~\cite{novakTyson}  
and  we present some preliminary experiments.
We adopt the following protocol: first we identify few BLTLc formulae 
characterising relevant aspects of the cell-cycle behaviour. We then   run our statistical 
verification tool to estimate the probability of the considered formulae.  
To assess the accuracy of the statistical procedure we 
compare the estimates obtained through our statistical model checker with 
the exact values calculated through numerical model checking, namely 
by means of the PRISM model checker~\cite{KNP09a}. 
As the state-space dimension corresponding to the original  
cell-cycle model is too large to be handled through numerical model checkers 
we consider a "scaled-down" version of the model for validating  
the statistical model checkin approach against the  numerical one.

Finally, for assessing the performances of the improved Wilson method  we 
compare the number of simulations performed by 
running every experiment twice: once with the initial point estimate $\alpha=1$, and then, 
following the original ``conservative" approach, with $\alpha=0.5$. 

The cell cycle is the a concatenation of biochemical and morphological events 
that lead to the reproduction (duplication) of a cell. 
The ``standard'' model considers a  loop of four phases:
\textit{G1}, growth and preparation of the chromosomes for replication;
\textit{S}, synthesis and duplication of DNA;
\textit{G2}, synthesis of significant protein for mitosis;
\textit{M},  mitosis, i.e., cell division.
The cell cycle is regulated by a network of biochemical reactions
centered around complexes of cyclin dependent kinases (Cdk's) and
their regulatory partners. Active complex Cdk/CycB 
induces cell cycle phase changes by activating or inhibiting target
molecules~\cite{Mor95,Mor06}.

\begin{figure}[!t] 
   \begin{center}
   \subfigure[]{\label{subfig:modelcartoon}}\includegraphics[scale=0.4]{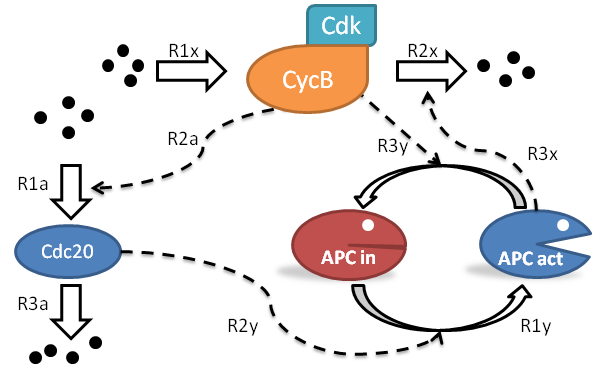} 
   \qquad 
   \subfigure[]{\label{subfig:simulation}}\includegraphics{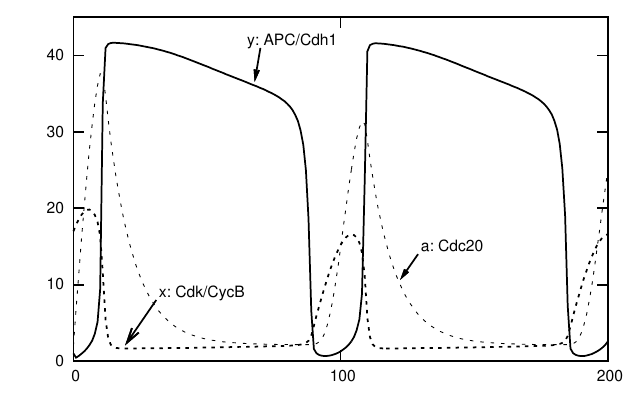} 
   \caption{Budding yeast cell-cycle cartoon and simulation}
   \label{fig:modelcartoon}
   \end{center}
\end{figure}

The simplified model we consider is sketched in Figure~\ref{subfig:modelcartoon}. 
It consists of three species, $x$ (Cdk/CycB complex), $y$ (activated APC/Cdh1 complex) 
and $a$ (activated Cdc20),
and nine molecular reactions listed in Table~\ref{tab:modelreactions} and with parameters 
in Table~\ref{tab:ParameterValuesCellCycleToyModel}.
 Complex $x$ is synthesized and degraded by reactions $R_{1x}$ and $R_{2x}$, respectively. 
Complex $y$ speeds up $x$ production by means of reaction $R_{3x}$. At the same time,
$x$ deactivates $y$ by $R_{3y}$. Also, $y$ turns active by itself with $R_{1y}$ and 
with the help of $a$ in reaction $R_{2y}$. 
Finally, $a$ is produced and consumed by  $R_{1a}$ and $R_{2a}$ and
regulated by $x$ in reaction $R_{2a}$.
Such system  behaves has a bistable switch with two stable states:
\text{G1} with low $x$ and high $y$, and \textit{S}/\textit{G2}/\textit{M} with high $x$ and low $y$,
as shown in the plot depicting a typical simulation outcome in Figure~\ref{subfig:simulation}.
\begin{table}[htbp]
\small
\begin{center}
\begin{tabular}{c|rcl||c|rcl||c|rcl|}
\hline
\multicolumn{4}{|c||}{Cdk/CycB} &  \multicolumn{4}{c}{APC/Cdh1} & \multicolumn{4}{||c|}{Cdc20}  \\  
\hline
\hline
\multicolumn{1}{|c|}{ $R_{1x}$ } & 
			$\emptyset$ &  $\xrightarrow{k_1 \alpha}$ &  $x$ 	&
$R_{1y}$ &						
			$y_{in}$ &  $\xrightarrow{k_3^*}$ & $y$	&
$R_{1a}$ & 
			$\emptyset$ &   $\xrightarrow{k'_5 \alpha}$ & $a$
\\
\multicolumn{1}{|c|}{ $R_{2x}$ } & 
			$x$ &  $\xrightarrow{k_2'}$ &   $\emptyset$	&		
$R_{2y}$ &						
			$y_{in} + a$ &  $\xrightarrow{k^{'''}_3}$ & $y + a$	&
$R_{2a}$ & 
			$x$ &   $\xrightarrow{k^*_5}$  & $x + a$
\\
\multicolumn{1}{|c|}{ $R_{3x}$ } & 
			$x + y$ &  $\xrightarrow{k^{''}_2 \alpha}$ &  $y$ 	&
$R_{3y}$ &						
			$x + y$ &  $\xrightarrow{k^{*}_4}$ & $x + y_{in}$	&			

$R_{3a}$ & 
			$a$ &   $\xrightarrow{k_6}$ & $\emptyset$
\\		
\hline
\end{tabular}
\end{center}
\normalsize
\caption{Budding yeast cell-cycle reactions}
\label{tab:modelreactions}
\end{table}

\begin{table}[htbp]
\small
	\begin{center}
		\begin{tabular}{|c|l|l|}
		\hline 
		{\bf Component} & {\bf Rate Constant} & {\bf  Dimensionless constants} \\
		\hline	
		Cdk/CycB & $k_1=0.04$, $k'_2=0.04$, $k''_2=1$, $k'''_2=1$  &  \\		 
		\cline{1-2}
		APC/Cdh1 & $k'_3=1$, $k''_3=10$, $k'_4=2$, $k_4=35$  &  $J_3 = 0.04, J_4 = 0.04$ \\
		& $k^{*}_3 = \frac{k_4  m \alpha x y}{J_4 + ( \alpha y) }$  
		$k^{'''}_3 = \frac{ k^{''}_3 \alpha a y_{in}   }{J_3 + ( \alpha y_{in}}$
				& $J_5 = 0.3$\\
		& $k^*_4 =  \frac{k'_3 y_{in}}{ J3 + ( \alpha y_{in})}$   &  $m = 0.80 $\\
		\cline{1-2}
		Cdc20 & $k'_5=0.005$, $k''_5=0.2$, $k_6=0.1$, $k_4=35$  & $\alpha = 0.00236012$ \\
		& $k^*_5 = \frac{k^{''}}{\alpha} / \frac{J_5}{m \alpha x} $ & \\
		& &\\
		\hline
		
		\end{tabular}
		\end{center}
		\normalsize
	\caption{Parameter Values: Cell Cycle toy model}
	\label{tab:ParameterValuesCellCycleToyModel}
\end{table}

\paragraph{Studying the role of Cdc20 in the \textit{S}/\textit{G2}/\textit{M} transition.}
We target the experiments of this section to the study of  the so-called \textit{S}/\textit{G2}/\textit{M} transition which begins in  states  
with low level of activated APC, high concentration of Cdk/CycB and (initially) low level of Cdc20. By looking at the topology of the network in Figure~\ref{subfig:modelcartoon}, and  
at the form of the corresponding equations (Table~\ref{tab:modelreactions}), it is evident that  $a$ (i.e. Cdc20) plays a fundamental part in the activation of $y$ hence in the  
controlling  the \textit{S}/\textit{G2}/\textit{M} transition. Specifically the progressive growth of $a$ results in the (initially slow) activation of $y$ which then, in turns, is responsible for the degradation of $x$. The influence of $a$ on $y$ can be studied 
through  BLTLc formulae of the following type:
$$
\phi_1\equiv (a \leq i)\ U\ (y\geq j), \hspace{4ex}
\phi_2\equiv (a \leq i)\ U^{\leq t} \ (y\geq j)
$$
Formula $\phi_1$ represents the possibility that $y$ grows above  threshold $j$ while $a$ does not exceed threshold $i$. Since $y$ gets abruptly activated only after $a$ 
has reached high concentration (see Figure~\ref{subfig:simulation}) then, for $i<j$ and $\delta=j-i$, we expect 
a low probability of $\phi_1$ for large  $\delta$, and a higher  probability of $\phi_1$ 
for high $i$ and small $\delta$\!\footnote{$\phi_2$ allows also 
to study the dependence on time of such an attitude.}\!.

\begin{figure}[!t] 
   \begin{center}
 \includegraphics[scale=0.6]{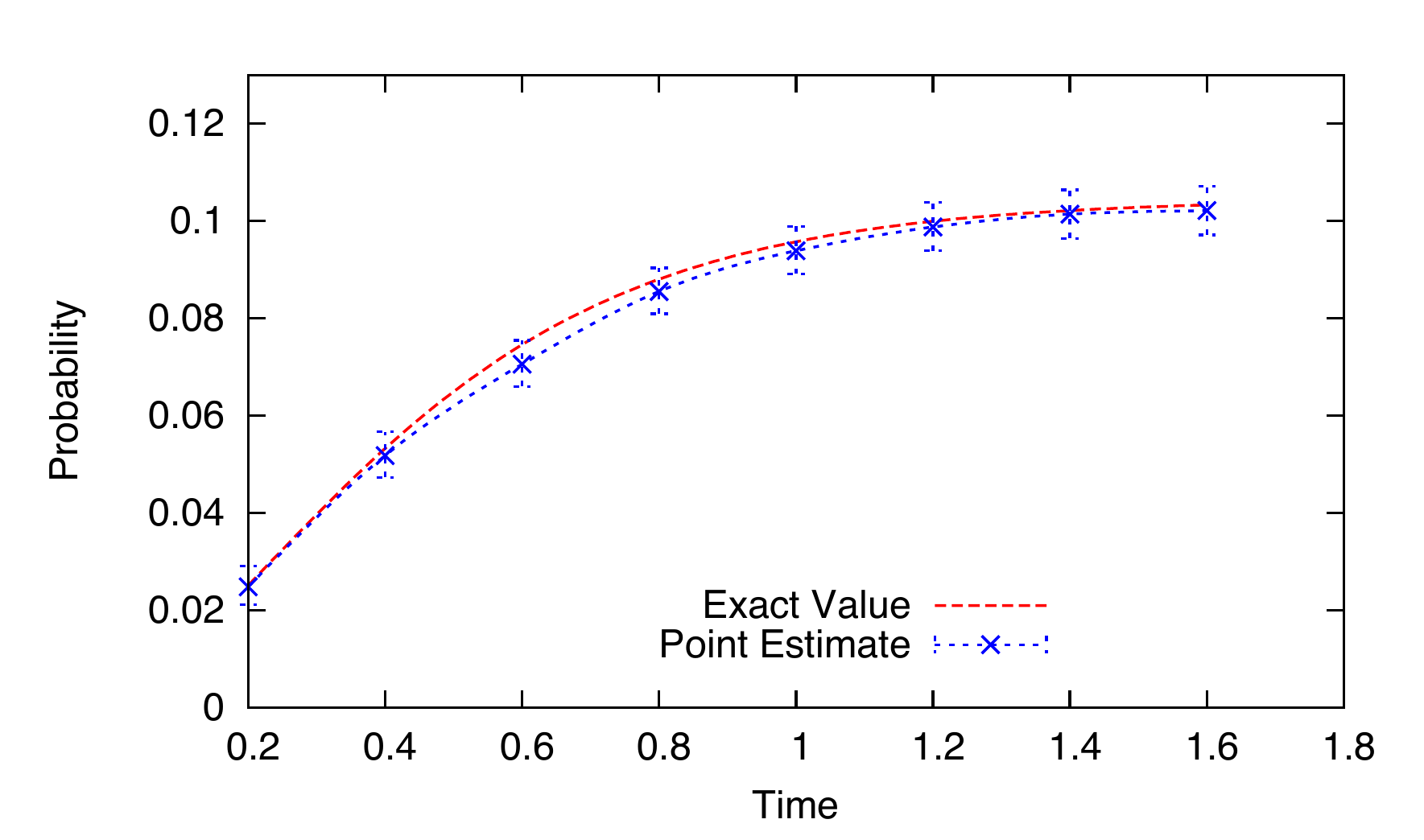} 
   \caption{Exact vs Estimated probability of the time-bounded Until formula: $(a \leq 4) U^{(0,t]} (y\geq 5)$, estimates with $99.99\%$ confidence  and $\epsilon=0.005$ semi-interval amplitude.}
   \label{fig:experiment1}
   \end{center}
\end{figure}

Figure~\ref{fig:experiment1}  compares  exact versus estimated probability measure 
for the time-bounded formula $\phi_2$, verified with respect to different time points ($t\in[0.2,1.6]$ step $0.2$). The (cross marked) point estimates (depicted in Figure~\ref{fig:experiment1} together with their confidence interval), have been calculated with $99.99\%$ confidence and $0.005$ interval semi-amplitude ($\epsilon=0.005$). The exact values computed with PRISM (red plot in Figure~\ref{fig:experiment1})
fall within the confidence interval of each point estimates, confirming the accuracy of the statistical 
verification method we have realized. 

Further data regarding application of statistical verification to  
formulae $\phi_1$ and $\phi_2$ (and variants) are reported in 
Table~\ref{table:experiments}. The calculated estimate ($\hat{p}$), 
confidence interval ($[L,U]$) corresponding to the chosen confidence level ($conf$) and interval width ($\epsilon$) are depicted together with the exact measure of probability $p$ (calculated with PRISM). For each experiment we also calculate the sample size ($N$) as well as the average path length $\overline{|\sigma|}$. 
Results in  Table~\ref{table:experiments} indicate a good accuracy of the obtained 
statistics and confirm the performance gain (i.e. in terms of sample size) of the sequential algorithm of Figure~\ref{fig:wilson} versus the conservative approach corresponding to $\alpha=0.5$. As argued before such  gain is greater with a larger distance of the estimate  
from the median.

\begin{table}[htdp]
\caption{Estimated probability of cell-cycle properties vs numerical evaluation.}
\small
\begin{center}
\begin{tabular}{|c|c|c|c|c|c|c|c|c|}
\hline
formula & $\hat{p}$ & $[L,U]$ & conf  & $\epsilon$  & $\alpha$ & $p$  & $N$ & $\overline{|\sigma|}$ \\
\hline 
$(a \leq 4)$ U $(y\geq 5)$ &  0.10384 & [0.080518,0.12700] & 99  & $0.025$  & 1 & 0.10458   & 1124 & 84.74 \\
\hline
$(a \leq 4)$ U $(y\geq 5)$ & 0,09554 & [0,081823,0,11128] & 99  & $0.025$  & 0.5 & 0.10458  & 2648 & 84,61 \\
\hline \hline

$(a \leq 20)$ U $(y\geq 35)$ &  0.00534 & [0.080518,0.12700] & 99  & $0.025$  & 1 & 0.00571  & 187 & 1274.65 \\
\hline
$(a \leq 20)$ U $(y\geq 35)$ & 0,00717 & [0,00401,0,01280] & 99  & $0.025$  & 0.5 & 0.00571  & 2648 & 1296,72 \\
\hline \hline

$(a \leq 40)$ U $(40\leq y\leq 42)$ &  0.62004 & [0.59472,0.64472] & 99  & $0.025$  & 1 & 0.6312  & 2495 & 4597.67 \\
\hline
$(a \leq 40)$ U $ (40\leq y\leq 42)$ & 0.63085 & [0.60064,0.65064 ] & 99  & $0.025$  & 0.5 & 0.6312  & 2648 & 4721,63 \\
\hline 
\end{tabular}
\end{center}
\label{table:experiments}
\end{table}%

\section{Conclusions}
\label{sec:Conclusion}
The stochastic modeling framework have been demonstrated  very important 
in systems biology. Unfortunately the complexity of living systems most often results in 
models which are so large that methods based on  
numerical approaches such as, for example, transient/steady-state analysis and/or 
\emph{exact} (stochastic) model checking are simply not feasible. For this reason, 
stochastic models of biological  systems are usually studied by means of simulation-based approaches. 
In this paper we have presented a statistical model checking approach targeted to the verification 
of biological models. 
A novel temporal logic language, namely the BLTLc language, has been introduced:  
it allows to formally capture  complex features of a biological system's dynamics. 
The methodology we proposed employes a statistical engine to estimate the probability 
of a BLTLc formula to hold true of a stochastic model $M$. Such estimates is obtained 
by \emph{on-the-fly} verification of the considered BTLTc formula $\phi$ against \emph{simulated executions} of 
the $M$. The resulting estimates is given by the frequency of the 
positive answers (i.e. the number of TRUEs resulting from the verification of $\phi$ against each simulation trace) out of the total of the simulated traces. 
The statistical engine we introduced is based on an efficient variant of the so-called Wilson 
score-interval method, which improves the performances, i.e. it requires a smaller  number of trajectories  in order to meet   the chosen confidence level, with respect to more popular statistical engines such as those based on the so-called Wald-interval method. 
We have implemented the proposed algorithms for analysis
and statistical testing as a prototype tool, which employs MPI technology to distribute verification engines.
The current implementation exploits multi-core processors, in particular, our tests have been  performed on a quad core machine
Intel Q9300 CPU with 4G of RAM under Widows XP. A cluster and a GRID version of our tool is under development, that will 
maximize the parallelism of the proposed methodology. Moreover we are planning a complete suite of performance
tests against tools with similar features, as those stated above. 

\textbf{Related work.} Techniques for the verification of temporal logic property against probabilistic/stochastic models, 
can be either \emph{exact} or \emph{approximate}. 
Exact approaches work by constructing a complete representation of a finite  state space model and, 
because of this, their application to  complex systems is unfeasible. 
PRISM~\cite{KNP09a} and MRMC~\cite{KatoenKZ_QEST05} are two popular probabilistic model checking tools that support 
both exact and approximated  CSL verification. 
Approximated verification can be one of two different types.
If the considered problem is to establishing whether
 the likelihood $p$ of a formula  is  $p \trianglelefteq b$ where $b\IN [0,1]$ is a threshold and $\trianglelefteq\{<,\leq,\geq,>\}$ (i.e.  \emph{model checking problem}) then the outcome of verification 
 is boolean and is determined based on Hypothesis testing. 
 On the other hand if the problem is one of determining an estimate for $p$ then this is 
 achieved through confidence interval based techniques. 
PRISM  approximated verification belong to the latter type:  the size of the 
 sample is determined statically as a function of the chosen level of confidence and the desired 
 approximation, rather than being calculated  iteratively as function of intermediate estimates, as is 
 the case with our method, whereas paths generation is controlled by on-the-fly checking of the 
 considered formula. Furthermore although PRISM has been recently added with support for (exact) probabilistic LTL model checking, at the best of our knowledge, it currently supports  statistical verification only for CSL (and not for LTL). 
The YMER~\cite{YKNP06} and MRMC tool, on other hand, features approximated (hypothesis testing based) model checking which uses on-the-fly verification of sampled path in order to decide whether 
 the probability of formula is below/above a threshold. 
The Monte Carlo Model Checker  MC2(PLTLc)~\cite{Donaldson08} computes a point estimate of a Probabilistic LTL logic (with numerical constraints) formula to hold of model. MC2(PLTLc) does not include any simulation engine but works \emph{offline} by taking a set of sampled trajectories  
 generated by any simulation or ODE solver software. Besides MC2(PLTLc) calculates also the 
probabilistic domain of satisfaction for any free variable of PLTLc formula. 
Finally the APMC tool~\cite{DBLP:conf/qest/HeraultLP06} features confidence interval based estimates 
of the probability of Probabilistic LTL and PCTL formulae to hold of either DTMC and CTMC models.\\[1ex] 
\noindent\textbf{Acknowledgments.}
The authors would like to thank Alida Palmisano for her valuable advises.

\bibliographystyle{eptcs}
\bibliography{PDMC09_EPTCS_BIBLIO}

\end{document}